\documentclass[superscriptaddress,aps,prx,reprint,nofootinbib,notitlepage,10pt,longbibliography]{revtex4-2}
\usepackage{graphicx} 
\usepackage{subcaption}
\usepackage{braket}
\usepackage{amsmath}
\usepackage{amssymb}
\usepackage{xcolor}
\usepackage{float} 
\usepackage{multirow}
\begin{document}

\title{Physics-Informed Neural Networks for Gate Design using Quantum Optimal Control}
\author{Sofiia Lauten}
\affiliation{%
    Department of Physics,
    University of Wisconsin -- Madison,
    Madison, WI 53706, USA
}%
\author{Matthew Otten}
\affiliation{%
    Department of Physics,
    University of Wisconsin -- Madison,
    Madison, WI 53706, USA
}%
\affiliation{%
    Department of Chemistry,
    University of Wisconsin -- Madison,
    Madison, WI 53706, USA
}%
\date{\today}

\begin{abstract}
Implementing quantum gates on quantum computers can require the application of carefully shaped pulses for high-fidelity operations. We explore the use of physics-informed neural networks (PINNs) for quantum optimal control to assess their usefulness in predicting such pulses. Our PINN is a feedforward neural network that utilizes an unsupervised learning approach, whose loss function includes terms that enforce the equations that govern the evolution of a quantum system, measure how close the learned unitary is to the target unitary operation, and ensure state normalization. We use a sinusoidal activation function and adopt variance-type weight initialization, tailored to our activation function. By analyzing the model's performance with important machine learning metrics, we demonstrate that the choice of our architecture is well-suited for this type of problem. We ensure that our network avoids the vanishing and exploding gradients with our relevant choices. We build two different PINNs, one based on the Schrödinger equation and another one based on the Lindblad equation. Our PINNs are able to discover high-fidelity two-qubit gate pulses for a variety of quantum operations, demonstrating its flexibility and robustness.
\end{abstract}

\maketitle

\section{Introduction}
 Quantum computers have promising potential to become valuable tools with advantageous applications for a variety of complex problems in simulation \cite{cao2019quantum,nguyen2024quantum}, optimization \cite{herman2023quantum}, and even machine learning \cite{biamonte2017quantum,otten2020quantum}. There are various architectures used for building a quantum computer, with their own benefits and drawbacks. A leading qubit architecture uses superconducting circuits \cite{devoret2013superconducting,mohseni2024build}. To perform computation on the qubits based on superconducting circuits, we have to implement quantum gates that manipulate the states of a qubit. Such gates can be realized by control pulses, generated by classical electronics outside the dilution refrigerator. The fidelity of gates can be affected by environmental noise, and if the pulses are properly optimized, the increased gate fidelity can drastically minimize noise effects and improve the reliability of operations \cite{matekole2022methods}.

Quantum optimal control (QOC) \cite {ansel2024introduction} can be used to search for optimized pulses to control the evolution of a quantum system. An optimization algorithm takes into consideration the type of task that a certain pulse has to achieve when one applies it. For example, it can create pulses that drive chemical reactions by breaking or forming molecular bonds \cite{dey2020controlling}, enhance the sensitivity of quantum sensors \cite{rembold2020introduction}, and improve precision in spectroscopic measurements \cite{keefer2021selective}. In this work, we focus on QOC's important ability to create pulses that are able to implement high-fidelity quantum gates that are essential for quantum computing. Manipulating an open quantum system, while using minimal resources, reaching the highest fidelities, and being robust against noise, is an important objective. 

From QOC theory, we can define control equations that will be responsible for designing and optimizing the needed pulses with appropriate constraints. We first have to understand if it is theoretically possible to drive a quantum system to perform a desired operation using the available control parameters. We can then choose the objective that we need to maximize at the final time and enforce the physical laws \cite{werschnik2007quantum}. It is also possible to explore experimental constraints by adding additional penalties. For example, the quantum speed limit must not be exceeded since it defines how quickly a state can evolve \cite{caneva2009optimal}. Bandwidth-limited pulses and their importance in helping experimental realization are also beneficial to be addressed \cite{rach2015dressing}. As another example, the addition of accounting for the strength of the control field (the amplitude) to be physically reasonable while respecting the hardware limits is another useful constraint \cite{machnes2018tunable}. After the theoretical setup of deriving the control equations that properly reach the goal is satisfied, we can explore ways of finding and optimizing the needed pulses. Methods such as chopped random-basis (CRAB) \cite{muller2022one}, Gradient Ascent Pulse Engineering (GRAPE) \cite{khaneja2005optimal}, and Krotov \cite{morzhin2019krotov} algorithmically search for solutions that satisfy these constraints, maximizing the fidelity and respecting physical limitations. In this work, we demonstrate the use of machine learning methods to solve QOC problems.

In QOC, we want to generate pulses to solve a desired task. Making the pulses obey the system’s equations of motion at every instant in time, while minimizing infidelity and the penalties (from experimental constraints), can be achieved by training a machine‐learning (ML) model, where we would have to tune weights to reduce a loss function. Just like in some other optimization methods (GRAPE), we can iterate through possible solutions, update the parameters using gradient information, and ultimately converge to an optimal set of parameters.

QOC can utilize machine learning (ML) in many ways. ML has proven to be a great tool to expand model space, allowing the control parameters to adopt a wider range of initial guesses \cite{mao2023machine}. When the model is not well known with insufficient equations to construct an accurate simulation for an optimization, techniques like Reinforcement Learning (RL) have been be introduced, incorporating a feedback loop to directly learn the optimal controls from the experimental data \cite{sivak2022model}. However, it is known that this closed-loop approach can be unable to generalize, and the learning task can become too complex. Another closed-loop technique to overcome some of RL's problems was introduced, using a supervised ML approach to learn the system dynamics from experimental data and then optimize the controls using the trained model \cite{genois2024quantum}.

Although the adaptability of controls to real-world environments is generally easier to achieve through closed-loop approaches, those tend to be more expensive to implement: one does not always have available experimental data. Instead, we turn our focus to unsupervised learning approaches. Even if we do not always know all the intricacies of the system in a non-model-free approach, we can still get a good enough approximate solution. For example, when tested with high noise presence, a GRAPE-based approach (an open-loop approach that does not rely on experimental data) outperformed an RL-based approach in \cite{fauquenot2024eo}.

\begin{figure*}
    \centering
    \includegraphics[width=0.99\linewidth]{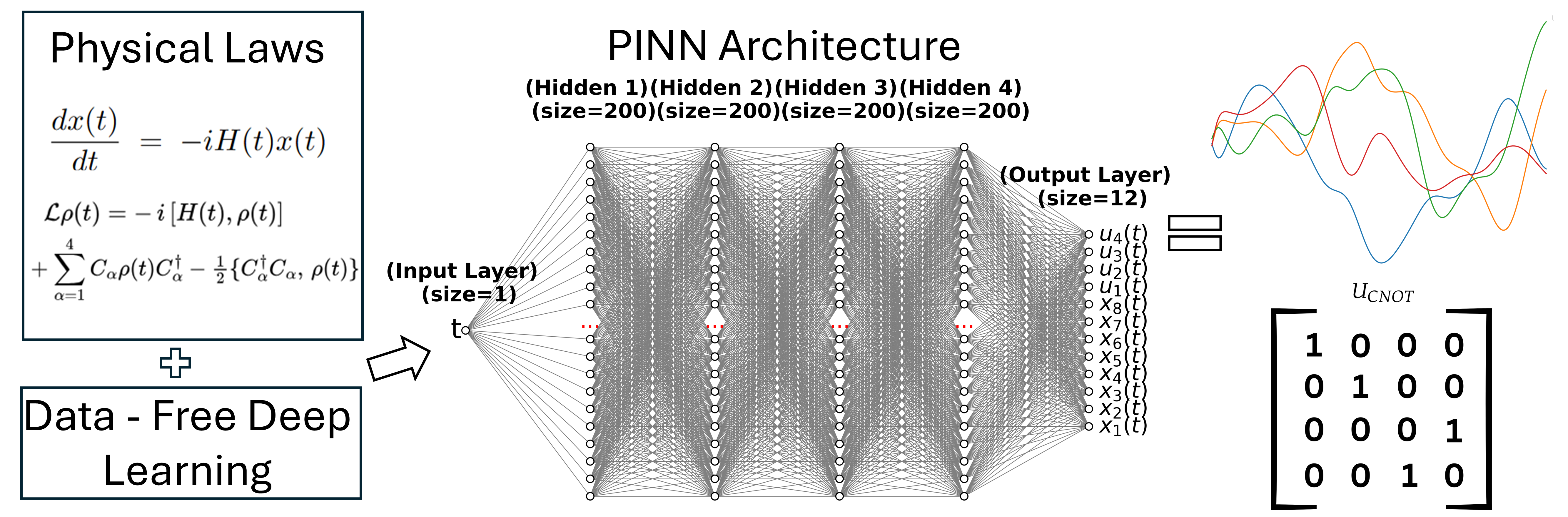}
    \caption{PINN overview and Neural Network architecture.}
    \label{fig:pinn_overview}
\end{figure*}

In this work, we explore the use of a Physics Informed Neural Network (PINN) for QOC. PINNs build physical laws directly into the training of the neural network, as shown in Figure~\ref{fig:pinn_overview}. Our exploration is inspired by \cite{norambuena2024physics}, where a PINN was developed to solve state-transfer problems. Their PINN finds smooth control functions for open quantum systems. The authors of that paper focus on preparing pulses to steer a two‑level system into a Gibbs mixed state, drive a three‑level system’s population from $\ket{1}$ to $\ket{2}$ via Stimulated Raman adiabatic passage (STIRAP) protocol, and additionally test their methods on a four‑level system’s population transfer. We expand these ideas to create a PINN that predicts smooth pulses to implement quantum gates for a two-qubit system. We first begin with a closed quantum system, where we model our PINN so that the entire unitary transformation matches a logical quantum gate. Then we expand our ideas to an open quantum system, based on the Lindblad master equation \cite{manzano2020short}. We find that our PINNs are able to achieve high-fidelity 
control sequences for a variety of gate targets, demonstrating the efficacy of 
the method.

\noindent

\section{PINN for implementing gates}
Our PINN is a feedforward neural network with an unsupervised learning approach. The key aspect is its loss function that enforces the physical laws that describe the needed dynamics. For our neural network architecture, the vector of times $t \in \mathbb{R}^{N \times 1}$ is our input. We have four fully connected hidden layers with 200 neurons each. It outputs a real-valued tensor \( y_{\text{out}}(t) \in \mathbb{R}^{N \times 12} \). The first eight values are grouped into two sets of four. The first set is interpreted as the real part and the other as the imaginary part of the complex-valued state network output, while the remaining four are used to define the control amplitudes $N_x(t) = x_{\text{out, real}}(t) + i\,x_{\text{out,imag}}(t), \quad N_u(t) = u_{\text{out}}(t)$. For all of the runs in both open and closed quantum system PINNs, we use $N=200$ time steps and $t=10$ (with units for time such that $\hbar = 1$). For optimization, we use Adaptive Moment Estimation (Adam) \cite{kingma2014adam}, with a learning rate of  $10^{-6}$, and $5000$ epochs (for almost all of the generated graphs unless specified otherwise) for our training. To add nonlinearity,  we apply a sinusoidal activation after the input layer, $\sin(\cdot)$, and after each hidden layer, $\sin(\omega_0\,\cdot)$ (the choice of activation is discussed later). The output layer is linear.

We model two qubits with a constant isotropic Heisenberg coupling \cite{nguyen2024programmable}, plus independently tunable transverse (x) and (y) control fields on each qubit \cite{hai2025scalable}. We first define Pauli X, Y, and Z matrices, along with the scaled identity operator,
\[
S_x = \sigma_x =
\begin{pmatrix}
0 & 1\\
1 & 0
\end{pmatrix},\qquad
S_y = \sigma_y =
\begin{pmatrix}
0 & -i\\
i & 0
\end{pmatrix},\qquad
\]
\vspace{0.05cm}
\[
S_z = \sigma_z =
\begin{pmatrix}
1 & 0\\
0 & -1
\end{pmatrix},\qquad
S_I = \frac12
\begin{pmatrix}
1 & 0\\
0 & 1
\end{pmatrix}.
\]
\\
We then define a drift Hamiltonian and the four control Hamiltonians,
\\
\begin{equation}
    H_d = \frac12\bigl(S_x\!\otimes\! S_x \;+\; S_y\!\otimes\! S_y \;+\; S_z\!\otimes\! S_z\bigr),
\end{equation}
with
\begin{equation}
    \begin{aligned}
        H_c^{(1)} &= S_x \!\otimes\! S_I, &\qquad
        H_c^{(2)} &= S_y \!\otimes\! S_I, \\[6pt]
        H_c^{(3)} &= S_I \!\otimes\! S_x, &\qquad
        H_c^{(4)} &= S_I \!\otimes\! S_y.
    \end{aligned}  
\end{equation}

\subsection{PINN Loss functions}
\subsubsection{The Loss Function for a closed quantum system}
As emphasized earlier, the loss function for a PINN is the key to the learning process: we subtract the predicted values by a neural network from the ones defined by physics equations. Here, we include the equations used in our method to compute our total loss function for our closed quantum system. 

First, we start out by defining a total Hamiltonian, 

\begin{equation}
H(t) \;=\; H_d \;+\; \sum_{j=1}^{n_{\mathrm{ctrl}}} u_j(t)\,H_c^{(j)},
 \label{eq:hamiltonian}
 \end{equation}
 where $H_d$ is the drift Hamiltonian, $H_c^{(j)}$ are the control Hamiltonians, and $u(t)=N_u(t)$ are the neural network predicted control amplitudes. This leads to the time-evolution operator,
\begin{equation}
      U_{t+\Delta t}
  \;=\;
  \exp\!\bigl(-\,i\,H(t)\,\Delta t\bigr)\,
  U_t .
\end{equation}

We now define the normalized state vector in terms of the neural network output, and we initially start at $x_0$,
\begin{equation}
x(t) = \frac{x_0 + \left(1 - e^{-t}\right) N_x(t)}{\left\| x_0 + \left(1 - e^{-t}\right) N_x(t) \right\|}.
\label{eq:norm}
\end{equation}
Such initialization scheme enforces smooth and bounded functions\cite{mattheakis2022hamiltonian}, where $N_x(t)$ are our neural network predicted state vectors. Then we define our key physics constraint for our PINN as the time-dependent Schrödinger equation,
\begin{equation}
\dot{x}(t)=\frac{dx(t)}{d t}
  \;=\;
  -iH(t)x(t).
\label{eq:schrodinger}
\end{equation}
We then take the time derivative using automatic differentiation (\texttt{torch.autograd.grad}) \cite{paszke2017automatic} of the state vector. Since our quantum state is complex-valued, we treat it as a pair of real-valued functions and apply automatic differentiation separately to the real and imaginary parts of the wavefunction. We first split $x(t)$,
\begin{equation}
x_{\text{real}}(t) = \Re[x(t)], 
\qquad 
x_{\text{imag}}(t) = \Im[x(t)].
\end{equation}

We then compute the derivatives for each component, for the $i$-th entry,
\begin{equation}
\dot{x}_{\text{real,i}}(t) = 
\frac{d}{d t}\,x_{\text{real},i}(t),
\quad
\dot{x}_{\text{imag,i}}(t) = 
\frac{d}{dt}\,x_{\text{imag},i}(t).
\end{equation}

To get the full complex time derivative, we recombine the real and imaginary parts together. This is a necessary step since most libraries still struggle to support automatic differentiation for complex functions \cite{krisha2024challenges}. A similar splitting technique is introduced in \cite{kramer2024tutorial} for complex differentiation,

\begin{equation}
\dot{x}_i(t) = \dot{x}_{\text{real,i}}(t) + 
i \,\dot{x}_{\text{imag,i}}(t).
\end{equation}
At each time step \(t_k\) for \(k = 1, \dots, 200\), the full derivative vector is:
\begin{equation}
\frac{dx(t_k)}{d t}=\dot{x}(t_k) = 
\begin{bmatrix}
\dot{x}_1(t_k) \\
\dot{x}_2(t_k) \\
\dot{x}_3(t_k) \\
\dot{x}_4(t_k)
\end{bmatrix}
\in \mathbb{C}^4 . 
\end{equation}
The complete time series $\dot{x}(t)_{predicted}$ is thus a tensor in \(\mathbb{C}^{200 \times 4}\), where each slice \(\dot{x}(t_k)\) is the complex derivative of the state vector at time $t_k$. We then minimize the difference between the learned time derivative, $\dot{x}(t)_{predicted}$, and the key physics constraint, equation \eqref{eq:schrodinger}, and therefore enforce the Schrödinger equation. To compute the $L_{model,closed}$, we calculate the squared Euclidean norm of the residual at each time step and average over all time steps. The Euclidean norm is appropriate for our learning here because our state is a vector,

\begin{equation}
L_{model,closed} = \frac{1}{N} \sum_{k=1}^{N} \left\| \dot{x}(t_k) - (-i H(t_k) x(t_k) )\right\|^2.
\end{equation}

Then we compute the fidelity where we compare the final learned propagator to the target propagator, where $d$ is the dimension of our system. We use process fidelity~\cite{gilchrist2005distance}, simplified for the unitary-unitary case. We define the infidelity as our second loss function $L_{\mathrm{fid}}$,

\begin{equation}
   \mathrm{F}(U_{\mathrm{targ}},U_t)  =  \frac{\bigl|\mathrm{Tr}\bigl(U_{\mathrm{targ}}\,U_t^\dagger\bigr)\bigr|}{d^2}^2,
\end{equation}

\begin{equation}
L_{\mathrm{fid, closed}}=1 - \mathrm{F}(U_{\mathrm{targ}},U_t).
\end{equation}

Lastly, we add the two loss terms together and get the final full loss function that the network can learn from. As a result, we can see that our loss function enforces the equations that govern the evolution of a quantum system and measures how close the learned unitary is to the target unitary operation,
\begin{equation}
L_{\mathrm{total, closed}}=L_{\mathrm{fid, closed}}+L_{model,closed}.
\end{equation}

\subsubsection{The Loss Function for an open quantum system}
For a $d$-dimensional system with the same Hamiltonian $H(t)$ from \eqref{eq:hamiltonian} and Lindblad (jump)
operators $\{C_\alpha\}$, the density matrix $\rho(t)$ evolves according to the
Lindblad master equation.
We consider a two-level system with ground state $\ket{g}$ and excited state $\ket{e}$.
At first, we define lowering and raising operators $\sigma_{ge} = \ket{g}\bra{e}, \quad \sigma_{eg} = \ket{e}\bra{g}.$
We include both spontaneous emission rate $\gamma_{\mathrm{em}}$, and absorption rate $\gamma_{\mathrm{abs}}$. 
The corresponding Lindblad jump operators are,

\begin{align*}
C_1 &= \sqrt{\gamma_{\mathrm{abs}}}\,(\sigma_{eg}\otimes I), &
C_2 &= \sqrt{\gamma_{\mathrm{em}}}\,(\sigma_{ge}\otimes I), \\
C_3 &= \sqrt{\gamma_{\mathrm{abs}}}\,(I \otimes \sigma_{eg}), &
C_4 &= \sqrt{\gamma_{\mathrm{em}}}\,(I \otimes \sigma_{ge}),
\end{align*}
where $I$ is the $2\times 2$ identity operator.  
Thus, the full set of collapse operators is,
\begin{equation}\label{eq:Cops}
\mathcal{C} \;=\; \{ C_1,\, C_2,\, C_3,\, C_4 \}.
\end{equation}

We use the same kind of state vector as in Schrödinger PINN, eq.~\eqref{eq:norm}, and construct our density matrix. We proceed to vectorize the density matrix because the Liouvillian can be written as a linear operator acting on vectors in  $\mathbb{C}^{d^2}$, instead of operators in  $\mathbb{C}^{d\times d}$,

\begin{equation}
\rho(t) = x(t)\,x(t)^\dagger \in \mathbb{C}^{4\times 4}, \quad
\vec{\rho}(t) = \mathrm{vec}\!\big(\rho(t)\big) \in \mathbb{C}^{16}.
\end{equation}
The corresponding Lindblad master equation with the $\mathcal{L}$, the superoperator, is therefore,
\begin{center}
\[
\frac{d}{dt}\rho(t) = \dot{\rho}(t)=\mathcal{L}\rho(t) = 
\]
\begin{equation} 
\mathcal{L}\rho(t) = -\,i\,[H(t),\rho(t)] + \sum_{\alpha=1}^{4}  C_\alpha \rho(t) C_\alpha^\dagger-\tfrac{1}{2}\{C_\alpha^\dagger C_\alpha,\,\rho(t)\}. 
\label{eq:lindblad} 
\end{equation}
\end{center}

We need to now define a proper fidelity for our open quantum system. We use a matrix representation of our open-system channel. To represent the full propagation, we perform a time-ordered product approximating the full channel. This approach was borrowed from \cite{zhang2025direct}, where they used the Lie–Trotter product formula to decompose the channel generated by a Liouvillian into a product of simpler exponentials. We use this same approach to approximate the evolution of our system by successive applications of short-time maps. $E_{\text{step}}$ is a short-time quantum channel, while $E_{\text{tot}}$ is the channel describing the open system evolution over the whole time window,
\begin{center}
   \[E_{\text{step}}(k) = e^{\Delta t \mathcal{L}_k}, \] 
\begin{equation}
E_{\text{tot}} \approx \prod_{k=0}^{N-1} E_{\text{step}}(k) = E_{\text{step}}({N-1})\cdots E_{\text{step}}(0),
\end{equation}
\end{center}
where $k=0,1,\dots,N-1$ is the discrete time-step index and $\Delta t = t/N$ is the step size. $\mathcal{L}_k$  is the Liouvillian superoperator constructed from the Hamiltonian $H(t)$ (with the current control amplitudes) 
and the collapse operators $\{C_\alpha\}$ at step $k$. Based on the \cite{gilchrist2009vectorization}, in the vectorized (Liouville space) representation, we can write a linear quantum channel 
as a matrix acting on the column-stacked 
form of an operator,
\begin{center}
\[
\mathrm{vec}(\mathcal{E}(A) = E_{\text{tot}}\,\mathrm{vec}(A),
\]
\begin{equation}
\label{eq:E-apply}
\mathcal{E}(A) = \mathrm{unvec}(E_{\text{tot}}\,\mathrm{vec}(A)\big),
\end{equation}
\end{center}
where $\mathrm{vec}(\cdot)$ is a column-stacking operation that maps a $d \times d$ matrix to a $d^2$-dimensional vector. The inverse operation $\mathrm{unvec}(\cdot)$ turns a $d^2$-dimensional back into a $d \times d$ matrix. Thus, $\mathrm{unvec}(\mathrm{vec}(A)) = A$. We are applying the channel to an operator $A$ by vectorizing $A$, acting with $E_{\text{tot}}$, and unvectorizing. Instead of any arbitrary A, for two qubits ($d=4$), we use the unitary basis
$\mathcal{U}_2=\{ \sigma_\mu\otimes\sigma_\nu : \mu,\nu\in\{0,x,y,z\}\}$,
with $\sigma_0=I$. It is orthogonal under the Hilbert–Schmidt inner product, satisfying
$\mathrm{tr}\!\big((\sigma_\mu\otimes\sigma_\nu)^\dagger
(\sigma_{\mu'}\otimes\sigma_{\nu'})\big) = d\,\delta_{\mu\mu'}\delta_{\nu\nu'}$ \cite{gilchrist2005distance}. 
Then, we are finally able to define process fidelity \cite{gilchrist2005distance},
\begin{equation}
F_{\mathrm{pro}}(\mathcal{E},U_{targ})= \frac{1}{d^3}
\sum_{P\in\mathcal{U}_2} \operatorname{tr}(U_{targ} P^\dagger U_{targ}^\dagger \mathcal{E}(P)),\quad d=4.
\label{eq:procfid}
\end{equation}

Now, it is possible to define the necessary loss functions for the open quantum system version of our PINN. We will start with our physics loss, where we minimize the difference between the $\dot{\rho}(t)$ and the right side of the equation \eqref{eq:lindblad}. Once again, for our time derivative, we use automatic differentiation, and since our $\rho(t)$ is complex-valued, we perform the same kind of splitting into real and imaginary, just like in the Schrödinger case,
\begin{equation}
\rho_{\text{real}}(t) = \Re[\rho(t)], 
\qquad 
\rho_{\text{imag}}(t) = \Im[\rho(t)],
\end{equation}

\begin{equation}
 \dot{\rho}_{\text{real,ij}}(t) = 
\frac{d}{d t}\,\rho_{\text{real},ij}(t),
\qquad
\dot{\rho}_{\text{imag,ij}}(t) = 
\frac{d}{dt}\,\rho_{\text{imag},ij}(t),
\end{equation}
     
\begin{equation}
    \dot{\rho}_{ij}(t) = \dot{\rho}_{\text{real,ij}}(t) + 
i \,\dot{\rho}_{\text{imag,ij}}(t).
\end{equation}
At each time step \( t_k \) for \( k = 1, \dots, 200 \), the full derivative matrix is:
\begin{equation}
    \dot{\rho}(t_k) = 
\begin{bmatrix}
\dot{\rho}_{11}(t_k) & \dot{\rho}_{12}(t_k) & \dot{\rho}_{13}(t_k) & \dot{\rho}_{14}(t_k) \\
\dot{\rho}_{21}(t_k) & \dot{\rho}_{22}(t_k) & \dot{\rho}_{23}(t_k) & \dot{\rho}_{24}(t_k) \\
\dot{\rho}_{31}(t_k) & \dot{\rho}_{32}(t_k) & \dot{\rho}_{33}(t_k) & \dot{\rho}_{34}(t_k) \\
\dot{\rho}_{41}(t_k) & \dot{\rho}_{42}(t_k) & \dot{\rho}_{43}(t_k) & \dot{\rho}_{44}(t_k)
\end{bmatrix}
\in \mathbb{C}^{4 \times 4}.
\end{equation}

\noindent
The full time series \(\dot{\rho}(t)\) is thus a tensor in \(\mathbb{C}^{200 \times 4 \times 4}\), with each slice \(\dot{\rho}(t_k)\) representing the derivative of the density matrix at time \(t_k\).
We compute $L_{model, open}$ by summing the squared Frobenius norms of the residuals. We choose to use a Frobenius norm since we are working with matrices now, instead of vectors. The Frobenius norm is defined as $\|A\|_F = \sqrt{tr[A^{T}A]}$ \cite{meurant1999computer}. As in the closed system case, we average over all $N$ time steps,
\begin{equation}
L_{model,open} \;=\; 
\frac{1}{N} \sum_{k=1}^N 
\left\| \dot{\rho}(t_k) - \mathcal{L}_k \rho(t_k) \right\|_F^2
\end{equation}

We can then define the infidelity using our process fidelity in equation \eqref{eq:procfid},
\begin{equation}
    L_{\text{fid,open}} = 1 -F_{\mathrm{pro}}(\mathcal{E},U).
\end{equation}

We include the trace loss, so that the trace of a density matrix is always equal to 1,
\begin{equation}
    L_{\text{trace}} = \frac{1}{N}\sum_{k=1}^N \Big(\mathrm{Tr}(\rho(t_k)) - 1\Big)^2.
\end{equation}

  The final total loss function for the open quantum system PINN can be defined as the following,
\begin{equation}
    L_{total,open}= L_{\text{fid,open}} + L_{\text{model,open}} + L_{\text{trace}}.
\end{equation}

\subsection{Weight Initialization choice}
A properly chosen initialization technique of weights (the numerical values that determine how strongly one neuron impacts another) can create faster convergence to a more correct solution and higher training accuracy \cite{narkhede2022review}. Additionally, a good weight initialization method could help avoid the vanishing \cite{wang2019reltanh} and exploding \cite{kanai2017preventing} gradients problems.

\begin{figure}
  \begin{subfigure}[b]{0.45\textwidth}   
        \centering\includegraphics[width=\linewidth]{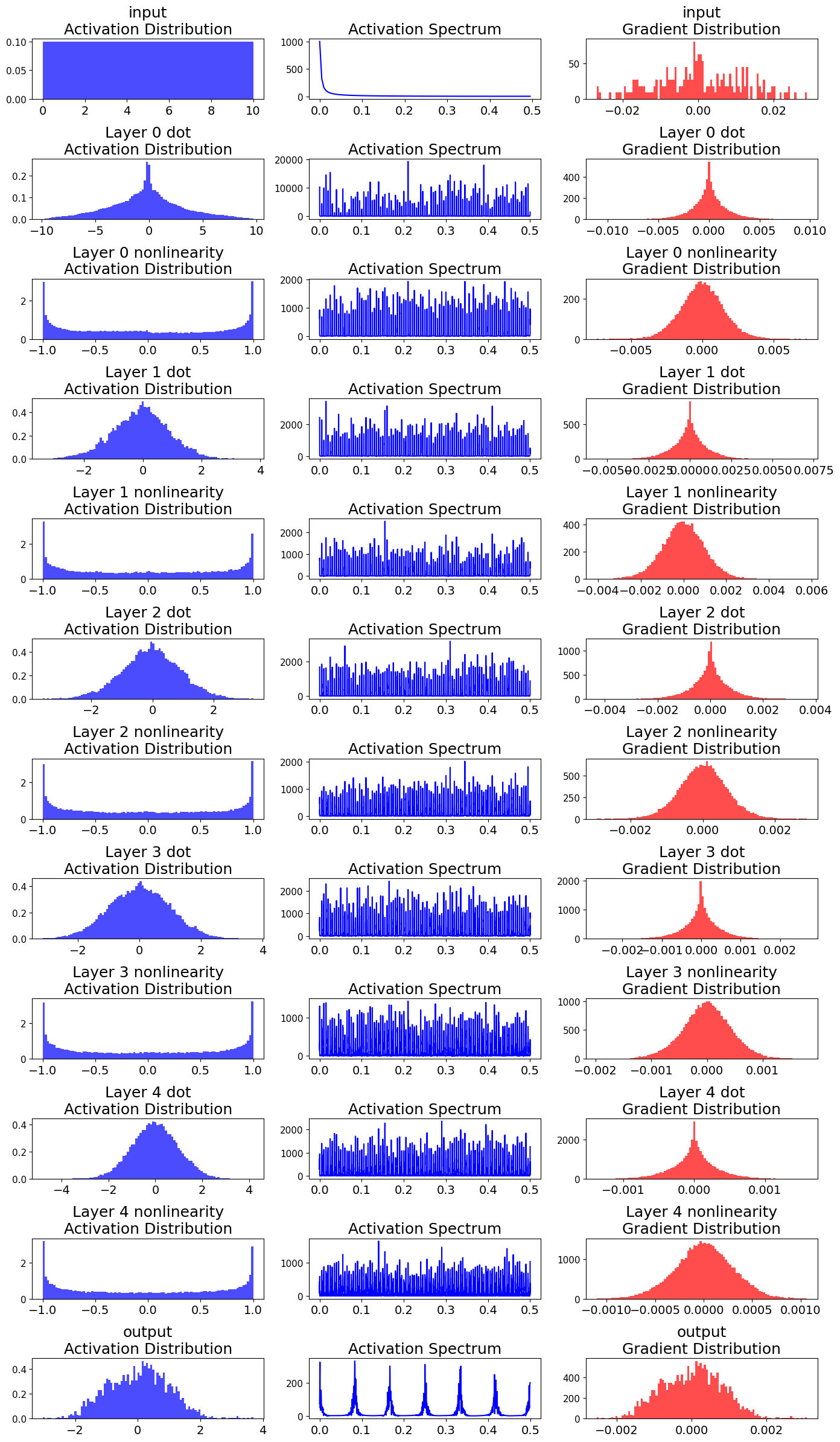}
  \end{subfigure}
  \caption{Visualization of activation distributions, gradient distributions, and activation spectrum across all the layers using the Schrödinger model trained with $U_{\mathrm{targ}} = \text{CNOT}$ and $\omega_0 = 1$, following the analysis of Ref.~\cite{sitzmann2020implicit}.}
  \label{fig:activation_distribution}
\end{figure}

We utilize the initialization scheme from Ref.~\cite{sitzmann2020implicit}, which also inspired our activation function, $\sin(\cdot)$. The network weights are sampled from a uniform random distribution over a specified range. It is a variance scaling type of weight initialization, more thoroughly analyzed in \cite{glorot2010understanding}. The initialization that was used for the input, hidden, and output layers is the following,

\begin{equation}
\begin{aligned}
W^{(\text{input})}_{ij} \sim \mathcal{U} \left( -\frac{1}{n_{\text{in}}}, \frac{1}{n_{\text{in}}} \right),\\ 
W^{(\text{hidden})}_{ij} \sim \mathcal{U} \left( -\frac{\sqrt{6/n_{\text{in}}}}{\omega_0}, \frac{\sqrt{6/n_{\text{in}}}}{\omega_0} \right),\\
W^{(\text{output})}_{ij} \sim \mathcal{U} \left( -\frac{\sqrt{6/n_{\text{in}}}}{\omega_0}, \frac{\sqrt{6/n_{\text{in}}}}{\omega_0} \right).
\end{aligned}
\end{equation}
In the results section, we discuss the influence of the $\omega_0$ values on our model's performance.
To evaluate the influence of the chosen initialization scheme, we performed the same visualization analysis on our PINN as in \cite{sitzmann2020implicit}. 

In Fig.~\ref{fig:activation_distribution}, we visualize the activation and gradient distributions as well as the activation spectrum across all the layers (using the Schrödinger model trained with $U_{\mathrm{targ}} = \text{CNOT}$ and $\omega_0 = 1$).
In our randomly initialized network (before training begins), distribution of activations and gradients stays approximately the same for each dot-product step, and for each nonlinearity step (after we apply the sinusoidal activation function), which indicates that the adopted weight initialization choices from \cite{sitzmann2020implicit}, similarly help to avoid vanishing and exploding gradients. The shape of activation distributions for each dot product is approximately Gaussian for most layers, centered around 0. For the nonlinearity step, the shape is sharply peaked near $-1$ and $+1$, with a flat region in between. The activation distributions are Laplacian-like for each dot product step and Gaussian-like for each nonlinearity step. The activation spectrum graph is spread across many frequencies, resembling a noisy spectrum. In the final output layer, the activation spectrum takes on sharp peaks at discrete frequencies.

\subsection{Sinusoidal Activation function relevance for PINN's}
The ability of a ML model to learn complex information (not just linear relationships in the data) is enabled by the addition of non-linear activation functions. Specifically for a PINN, it is demonstrated in \cite{wang2023learning} that these networks are generally sensitive to the choice of activation functions. Since, in our case, we do not have any previous data to learn from (we are just relying on the equations), it would be immensely helpful to have an activation function that fits the nature of the problem better (for example, periodic), allowing the network to yield better predictions. 

\begin{figure}
  \centering
  \begin{subfigure}{\linewidth}
    \centering
    \includegraphics[width=0.7\linewidth]{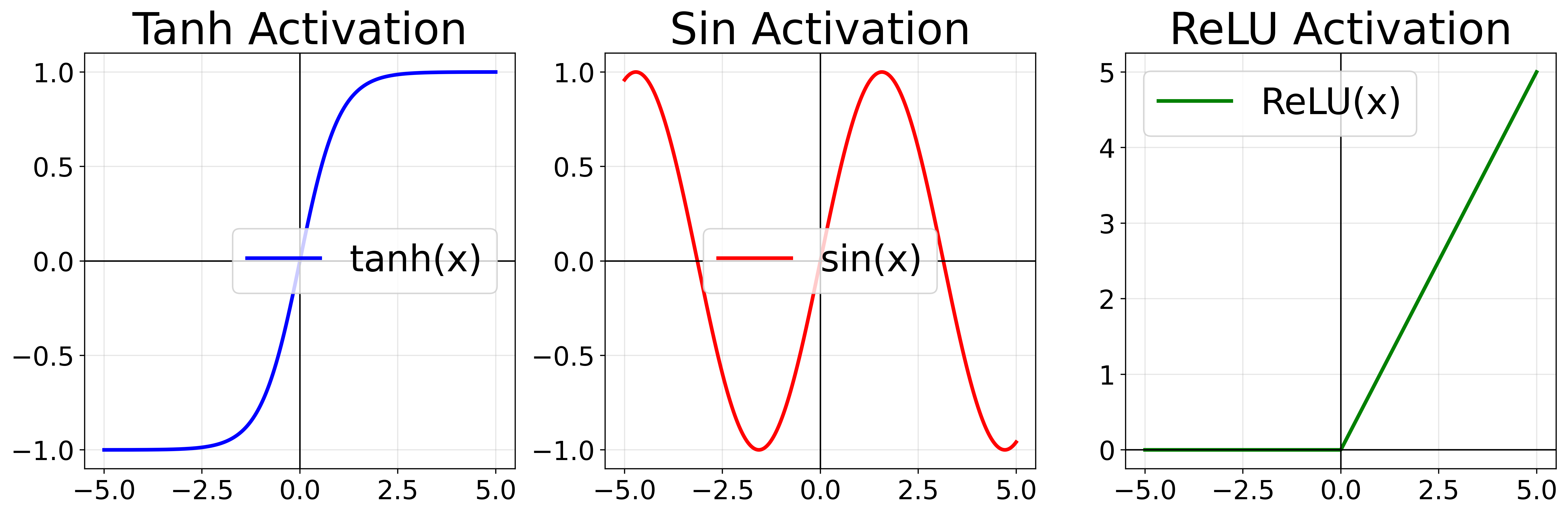}
  \end{subfigure}

  \begin{subfigure}{\linewidth}
    \centering
    \includegraphics[width=0.7\linewidth]{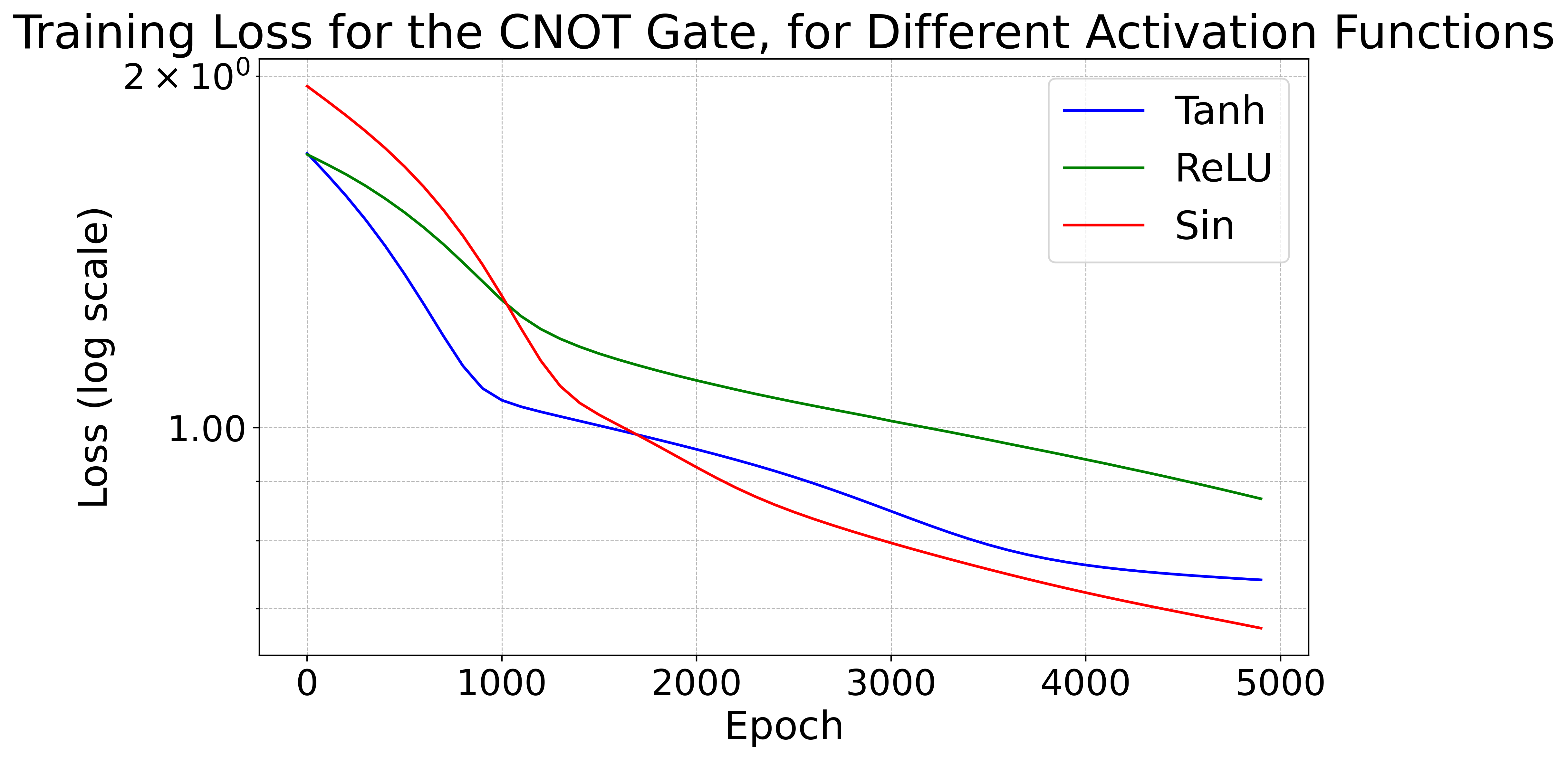}
  \end{subfigure}

  \caption{Top: Different Activation Functions. Bottom: Fidelity vs Epochs for different activation functions for the Schrödinger model trained on $U_{CNOT}$.}
  \label{fig:activation_comparison}
\end{figure}

We chose to use a sinusoidal activation function for our PINN since it has been shown in \cite{zeng2022adaptive} that networks utilizing the $\sin(\cdot)$ activation function can successfully solve partial differential equations. To demonstrate that our choice of activation function works better than conventional approaches like $\tanh(\cdot)$ and $\mathrm{ReLU}(\cdot)$, we have visualized the training of our model using all of the aforementioned activation functions and plotted the loss at each epoch throughout the entire training of our model (Fig.~\ref{fig:activation_comparison}).

The underperformance of the $\mathrm{ReLU}(\cdot)$ activation function is most likely coming from the fact that its second-order derivative is zero, as highlighted in \cite{wang2024artificial}. The authors of that paper point out that in PINNs, a solution must satisfy the differential structure of the governing equations, so activation functions with discontinuous (or zero) higher-order derivatives do not work. We can see that $\tanh(\cdot)$ also performs worse than $\sin(\cdot)$. Ref.~\cite{faroughi2023physics}, which proposes the use of PINNs for the solute transport problems, similarly shows the superiority of the sinusoidal activation function when compared to $\tanh(\cdot)$, with the training being faster and results being more accurate when trained with $\sin(\cdot)$. As \cite{faroughi2023physics} points out, $\sin(\cdot)$ is naturally suited for learning wavy, repeating behaviors, and training with it is smoother and faster because itself and its derivative, $\cos(\cdot)$, are closely related.
There are numerous other works that discover the benefits of the sinusoidal activation function, such as Ref.~\cite{mattheakis2022hamiltonian}, where using $\sin(\cdot)$ improves their Hamiltonian neural network's ability to learn periodic or chaotic trajectories, resulting in faster convergence.

\section{Results}
\begin{figure}[h!]
\centering
  \begin{subfigure}[b]{0.42\textwidth}
  \centering
    \includegraphics[width=0.9\linewidth]{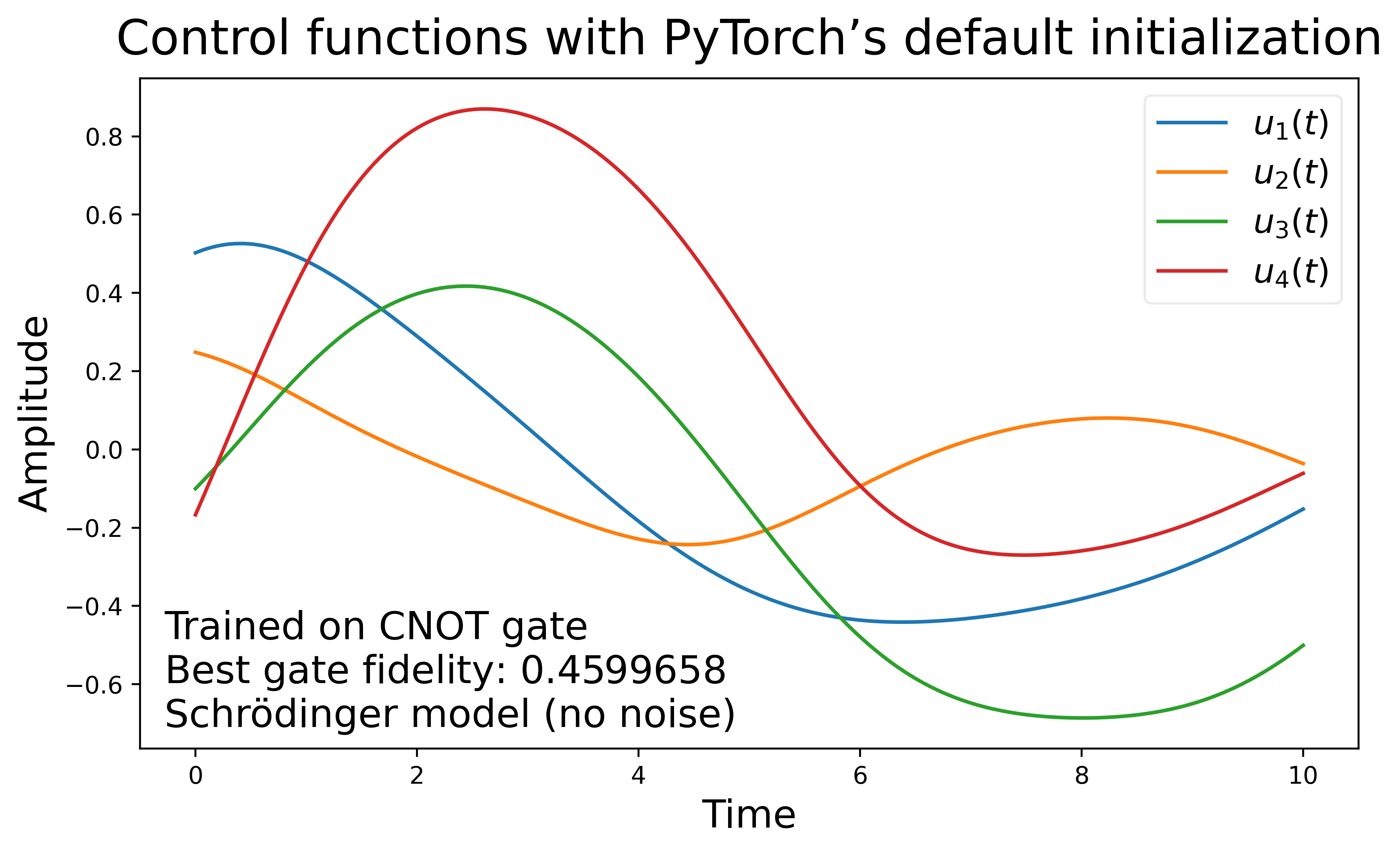}
  \end{subfigure}
   \begin{subfigure}[b]{0.39\textwidth}
   \centering
   
       \includegraphics[width=0.95\linewidth]{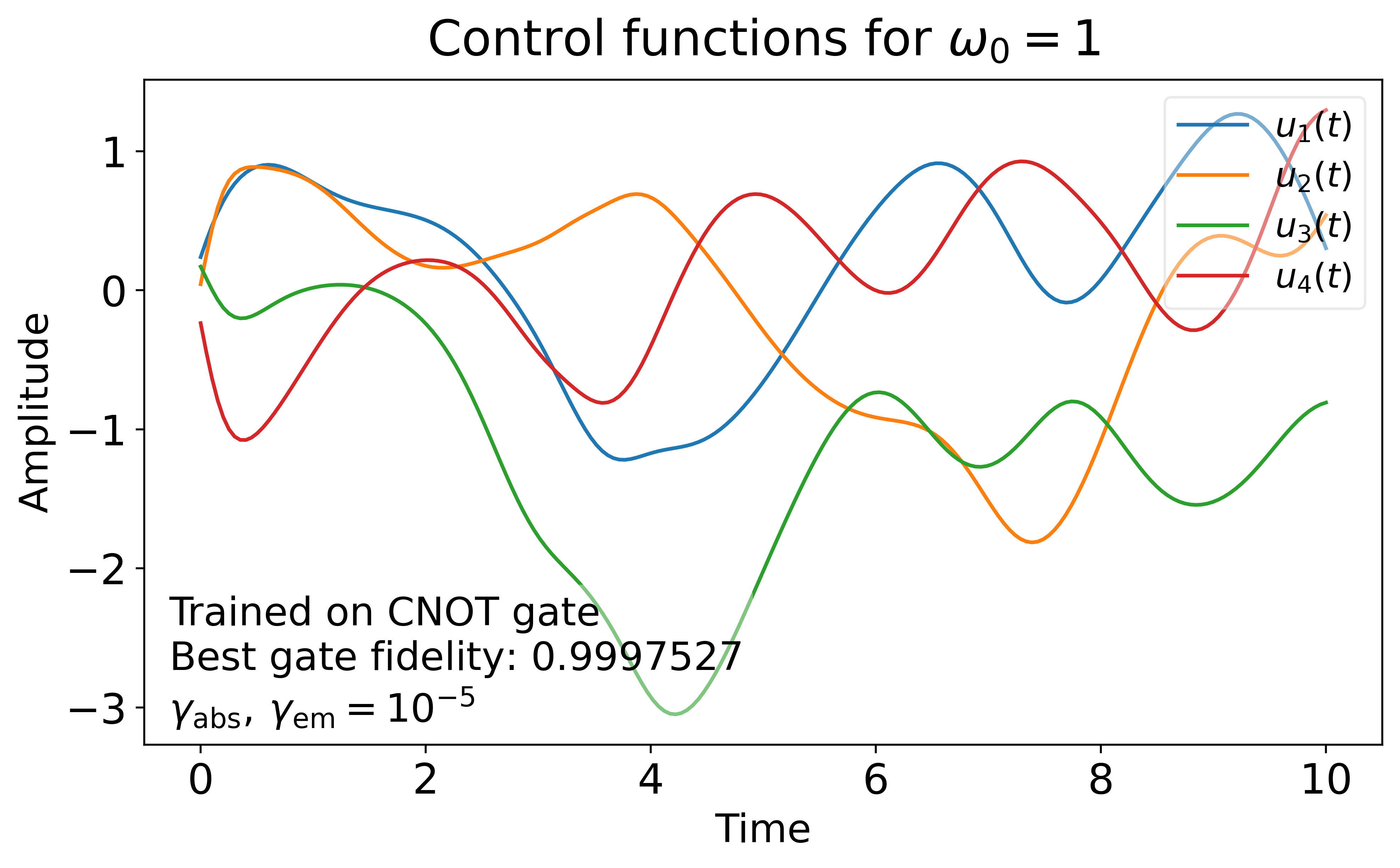}
   \end{subfigure}
    \begin{subfigure}[b]{0.4\textwidth}
   \centering
       \includegraphics[width=0.95\linewidth]{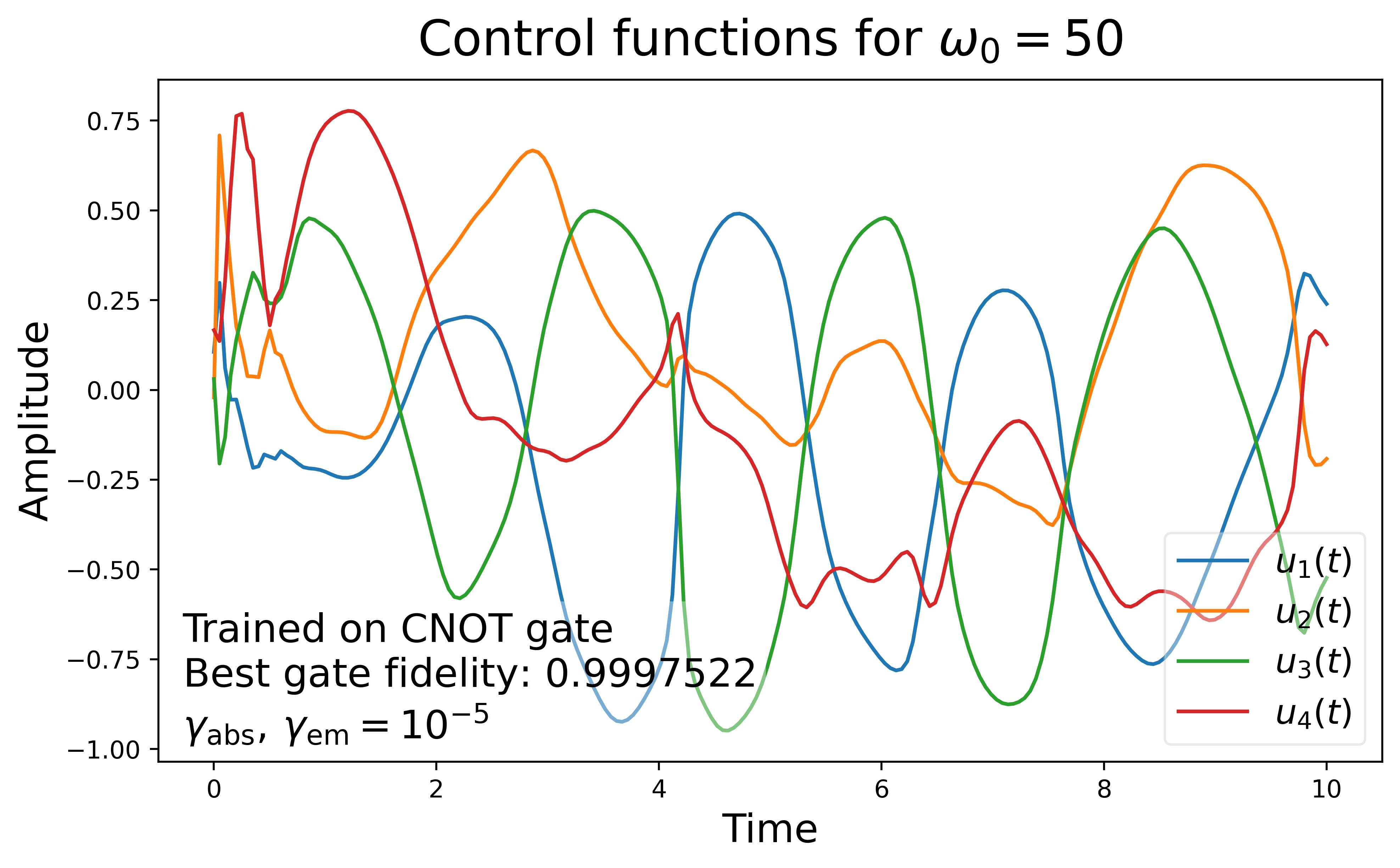}
   \end{subfigure}
    \begin{subfigure}[b]{0.4\textwidth}
   \centering
       \includegraphics[width=0.95\linewidth]{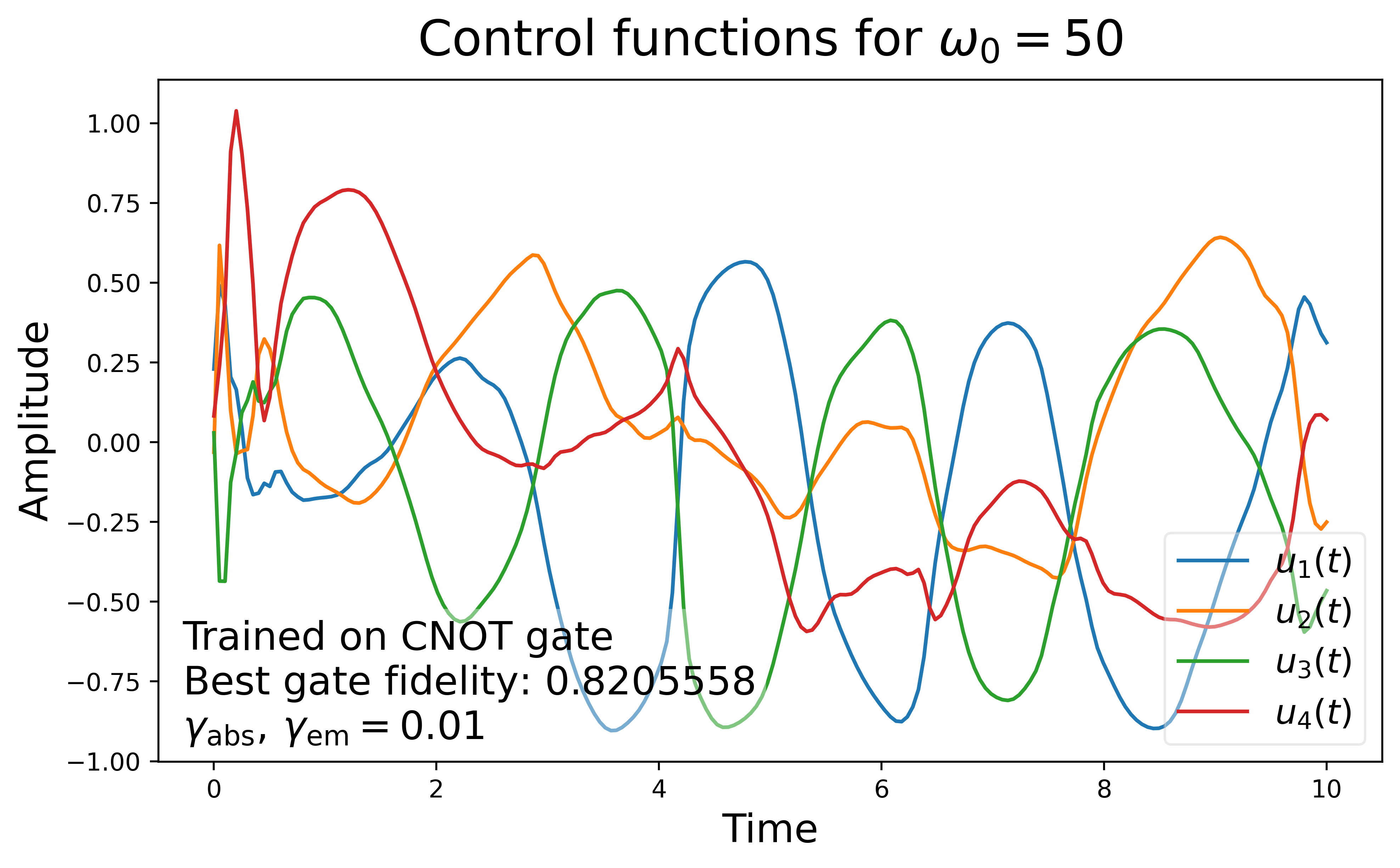}
   \end{subfigure}
   \begin{subfigure}[b]{0.38\textwidth}
   \centering
       \includegraphics[width=0.95\linewidth]{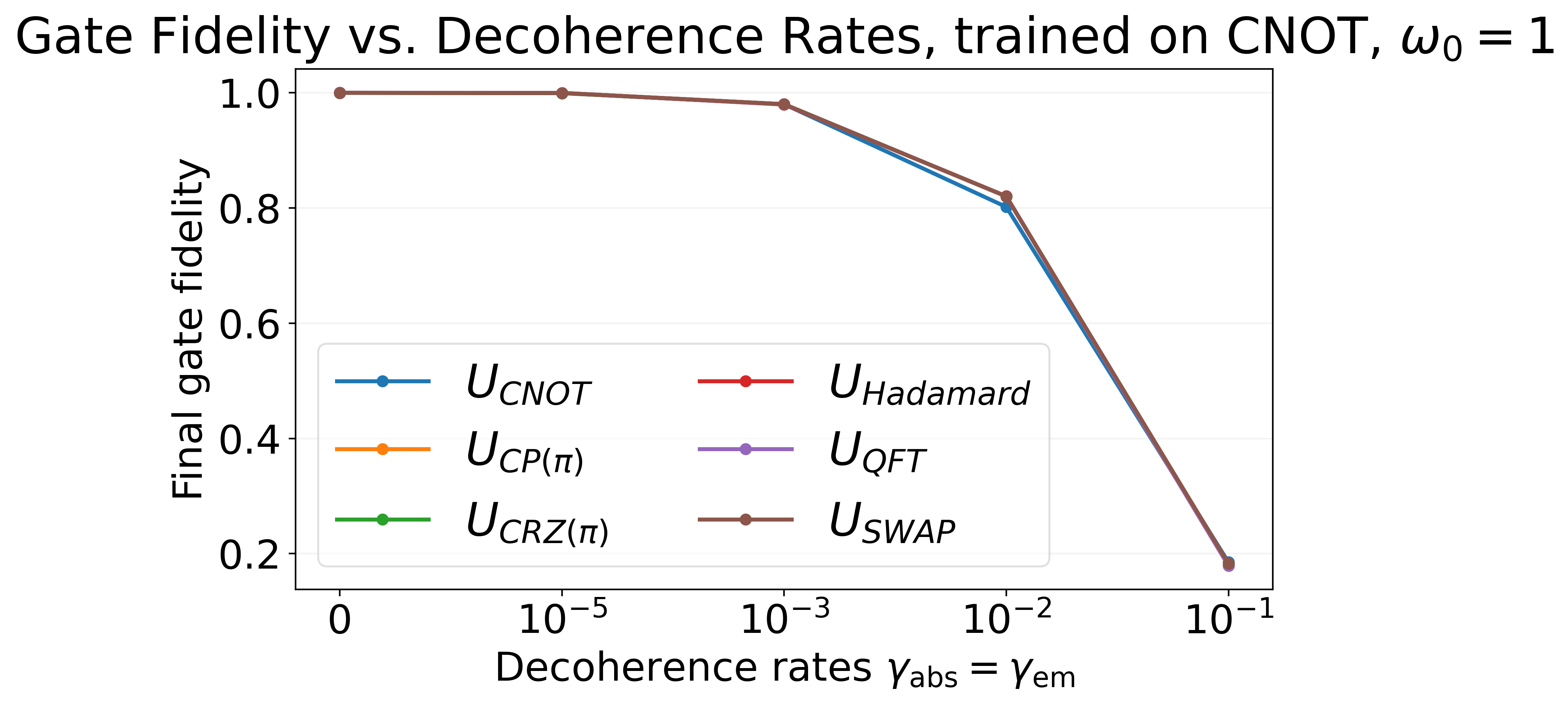}
   \end{subfigure}
   \caption{Control amplitudes for models trained with custom initialization for different $\omega_0$ vs a model trained with weight initialization using PyTorch’s default initialization for nn.Linear (first four graphs from the top). Maximum epochs were set to 4000. Decoherence rates influence on final gate fidelities (last graph).}
   \label{fig:control_amps}
\end{figure}

To choose an optimal $\omega_0$ value for our neural network's weight initialization, we tested different $\omega_0$ values to see the influence on the final pulse outputs. In Fig.~\ref{fig:control_amps}, we visualize the different control functions for three runs with different weight initialization cases for the Lindbladian model, trained on $U_{CNOT}$, and also a run where all linear layers were initialized using PyTorch's default initialization scheme for nn.Linear for the Schrödinger model.

Increasing $\omega_0$ can sometimes slightly reduce the amplitudes of the control functions  $u_{j}(t)$, but still allows the model to achieve similarly high fidelities. When increasing the $\omega_0$ value for the Lindbladian model with low noise rates of $\gamma_{\mathrm{abs}}, \gamma_{\mathrm{em}} = 10^{-5}$, we can see that the $\omega_0 =1$ produced smoother, sinuisoidal pulses, unlike $\omega_0=50$, which had sharper and rougher pulses (in general with much higher oscillations) in some places. Similar performance in pulse shapes is achieved in a noisier Lindbladian model, $\gamma_{\mathrm{abs}}, \gamma_{\mathrm{em}} = 0.01$, indicating that the main reason for rigid pulses is the higher $\omega_0$ value, not the higher noise rate. Such highly oscillatory and sharper pulses contain lots of high-frequency Fourier content, and in general, are much harder to experimentally realize due to constraints of the control electronics, like the pulse-modulation bandwidth \cite{krantz2019quantum}. Additionally, sharp discontinuities are physically forbidden since an instantaneous increase in voltage or current would demand unbounded power \cite{hartnagel2023fundamentals}. For our case, it is most optimal to therefore use $\omega_0 = 1$, but still keep all of the other weight initialization parts similar to what was done in \cite{sitzmann2020implicit}. 

\begin{figure}
 \includegraphics[width=\linewidth]{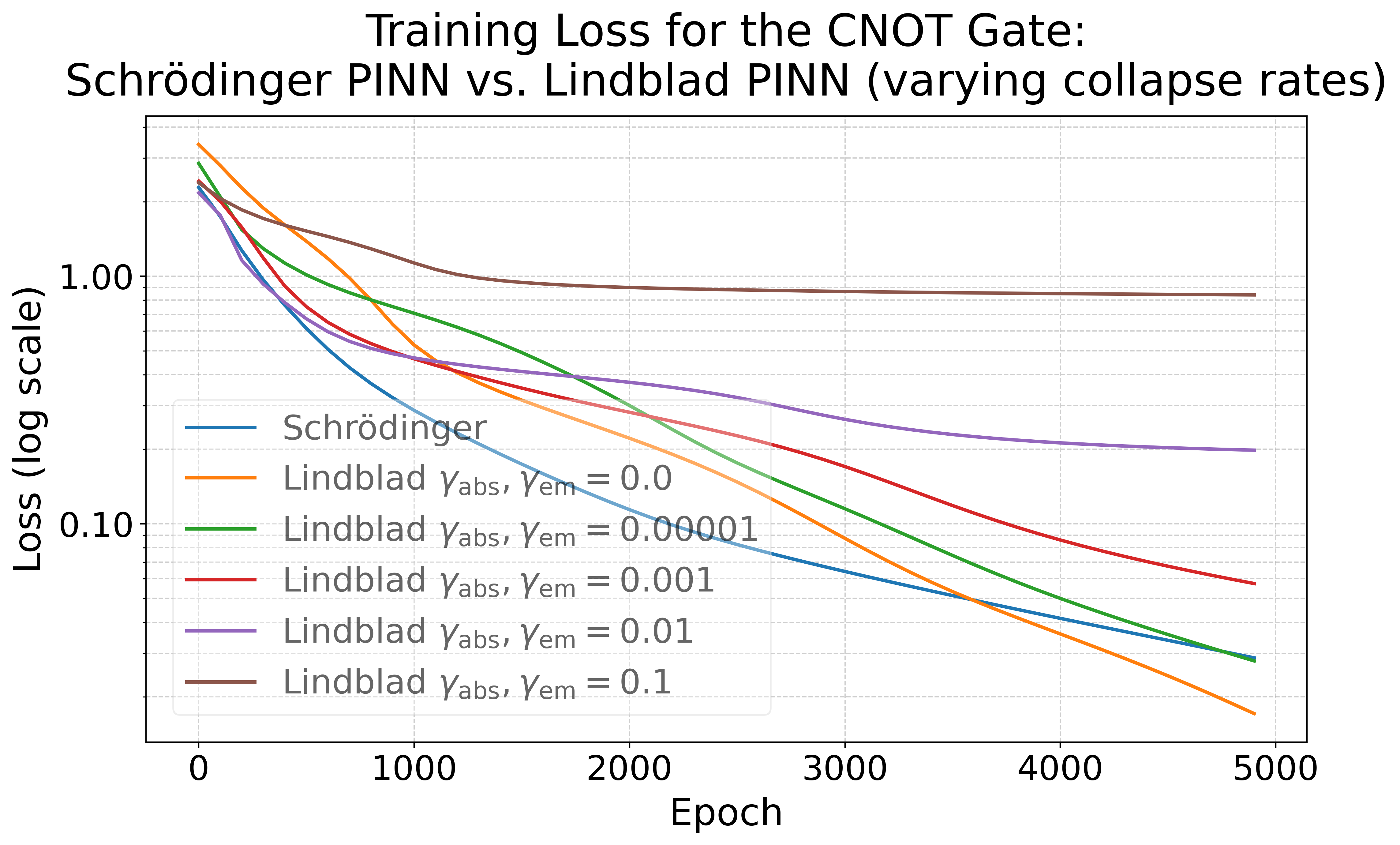}
  \caption{Training Loss for models trained on $U_{CNOT}$ Gate over 5000 epochs for runs of Schrödinger PINN and Lindblad PINN (with different rates of the collapse operators).}
  \label{fig:training_loss}
\end{figure}

When PyTorch's default weight initialization for nn.Linear was used even in the noiseless environment; the gate fidelity at the final epoch had only reached $0.4599658$, so it significantly underperformed compared to our custom weight initialization models. Additionally, changing the $\omega_0$ value did not have any influence on improving gate fidelities for the higher noise rate versions of Lindbald PINN ($\gamma_{abs}=\gamma_{em}=0.01$ and $\gamma_{abs}=\gamma_{em}=0.1$). The model performs in a similarly bad way for higher noise rates, no matter what gate it is trained on, which is shown in the Gate Fidelity vs Decoherence Rates graph in Fig.~\ref{fig:control_amps}. In that graph, each corresponding PINN model was trained on each gate five times for the different decoherence rates (consisting of the same rates that are shown in Fig.~\ref{fig:training_loss}), totaling 30 different Lindblad models overall. 

To thoroughly test our Schrödinger PINN model, we have trained it on different target gates, $U_{\mathrm{targ}}$. We target $U_{\mathrm{CNOT}}$, a standard choice for a two-qubit gate,  that, in conjunction with single-qubit gates, leads to universality \cite{nielsen2001quantum}. We also train on $U_{\mathrm{CP}(\theta)}$ \cite{qiskitCPhaseGate} and $U_{\mathrm{CRZ}(\theta)}$ \cite{qiskitCRZGate}; two other common two-qubit entangling gates. We test a two-qubit Hadamard 
$U_{\mathrm{Hadamard}} = H \otimes H$, which creates product superpositions and is a tensor product of one-qubit gates \cite{nielsen2001quantum}; this evaluates whether our PINN can properly optimize simultaneous single-qubit controls on both qubits. We then test our PINN on $U_{\mathrm{SWAP}}$, which is a gate that helps to bring information from non-adjacent qubits together \cite{wille2014optimal}. We additionally test the PINN's ability to generate the two-qubit quantum Fourier transform $U_{\mathrm{QFT}}$, which is a combination of both single-qubit and two-qubit gates.

As shown in Table~\ref{tab:fids}, we have successfully reached high fidelities for all six two-qubit gates tested (after training the model for 5000 epochs and setting  $\omega_0 = 1$ in the weight initialization). To ensure meaningful testing of each target gate, each time we trained on a new $U_{targ}$, we initialized the quantum state vector from equation \eqref{eq:norm} with an appropriate $x_0$ so the specific action of the gate can be properly highlighted. This way, we are able to correctly evaluate the gate’s operational effect in the learned dynamics. The approximate time for training (on a commodity computer) was about 10 minutes for all of the gates. 
\begin{center}
\begin{tabular}{|c|c|}
\hline
\textbf{Gate} & \textbf{Final Gate fidelity} \\
\hline
$U_{QFT}$ & 0.9995733499526978 \\
\hline
$U_{Hadamard}$ & 0.9989193081855774 \\
\hline
$U_{CNOT}$ & 0.9998568892478943 \\
\hline
$U_{CRZ({\pi})}$ & 0.9999414682388306 \\
\hline
$U_{CP({\pi})}$ & 0.9998995661735535 \\
\hline
$U_{SWAP}$ & 0.9997590780258179 \\
\hline

\end{tabular}
\captionof{table}{Final fidelities for different gates for the Schrödinger PINN.}\label{tab:fids}
\end{center}

Additionally, we decided to further test our PINN-predicted pulses by using the final-learned control amplitudes for some gates to initialize one existing and well-established algorithm, CRAB, and evaluating CRAB's performance. For all of the runs, CRAB that was initialized with PINN pulses performed as well (if not a little better for some gates) as CRAB on its own, indicating that our solutions work well. We used the generic public QuTiP notebook implementation of CRAB for the two-qubit QFT gate (from which we also adopted the basic physics definitions of our drift and control Hamiltonians) \cite{qutipCRABQFT}.

For the Lindbladian version of our model, we also trained it on each gate, for each different value of $\gamma_{\mathrm{abs}}$ and $\gamma_{\mathrm{em}}$ (for 5000 epochs). We used our custom weight initialization with $\omega_0 = 1$. Our final fidelities from gate to gate were very similar, which is demonstrated in the last graph of Fig,~\ref{fig:control_amps}. The approximate time for training was about 22 minutes for all of the runs. 

To visualize the difference in training between the different rates of collapse operators for the open quantum system PINN and the closed one, we show the overall loss vs epochs for the $U_{CNOT}$ in Fig.~\ref{fig:training_loss}.
As expected, the Schrödinger PINN and the Lindblad PINN with zero (or very small) noise rates achieve the lowest training loss throughout all epochs (our Lindblad $\gamma_{abs},\gamma_{em}= 0.0$ model achieved an even lower loss value than Schrödinger PINN at very late epochs, which is likely do to choice of hyperparameters). With this, we confirm that our Lindblad PINN with zero collapse rates effectively simplifies to a closed-system evolution. Overall, the initial loss values and convergence trends vary noticeably. As the collapse rates increase, the losses converge more slowly, which corresponds to the difficulty of learning in the presence of decoherence and dissipation.

It is known that setting the collapse operators to zero effectively reduces the Lindbladian dynamics to the standard Schrödinger dynamics. To further verify the physical consistency of our PINN implementation, we conducted a validation check under this condition. Specifically, we trained the Lindbladian version of our model while setting the collapse operators to zero, extracted the learned control functions, and used them to evolve the system using QuTiP’s mesolve (a master equation solver) \cite{lambert2024qutip}. We did so by inserting the learned control functions into a time-dependent Hamiltonian, which then becomes one of the parameters we input into mesolve. It was necessary for us to use interpolation for the control functions before we inserted them into the Hamiltonian because QuTiP’s mesolve requires time-dependent coefficients to be defined as continuous functions rather than discrete samples. We specifically chose a cubic spline, as was done in \cite{ernst2025memory}, for smoothness and experimentally accurate control pulses. In \cite{li2022pulse}, a cubic spline was shown to specifically be a better addition if one plans on using QuTiP solvers. We repeated the same procedure for the Schrödinger version of our model. After evolution, we compared the final density matrices produced by both approaches. As shown in Table~\ref{tab:fids_dm}, the high fidelity values between the two final states confirm that our PINN framework correctly recovers closed-system dynamics when dissipation is absent.
\begin{center}
\begin{tabular}{|c|c|}
\hline
\textbf{Gate} & \textbf{Fidelity of Final Density Matrices} \\
\hline
$U_{QFT}$ & 0.9984088682348283 \\
\hline
$U_{Hadamard}$ & 0.9941456388008292 \\
\hline
$U_{CNOT}$ & 0.9968657487935076 \\
\hline
$CRZ({\pi})$ & 0.999231166866835 \\
\hline
$U_{CP({\pi})}$ & 0.9993394261008633 \\
\hline
$U_{SWAP}$ & 0.9996757404438216 \\
\hline

\end{tabular}
\captionof{table}{Fidelities between the final density matrix after evolving the system with mesolve that used Lindblad learned control functions (when setting collapse operators to zero) in its Hamiltonian vs the final density matrix after evolving the system with mesolve that used the Schrödinger learned control functions in its Hamiltonian.}\label{tab:fids_dm}
\end{center}

\begin{figure}
  \begin{subfigure}[b]{0.45\linewidth}   
        \centering\includegraphics[width=\linewidth]{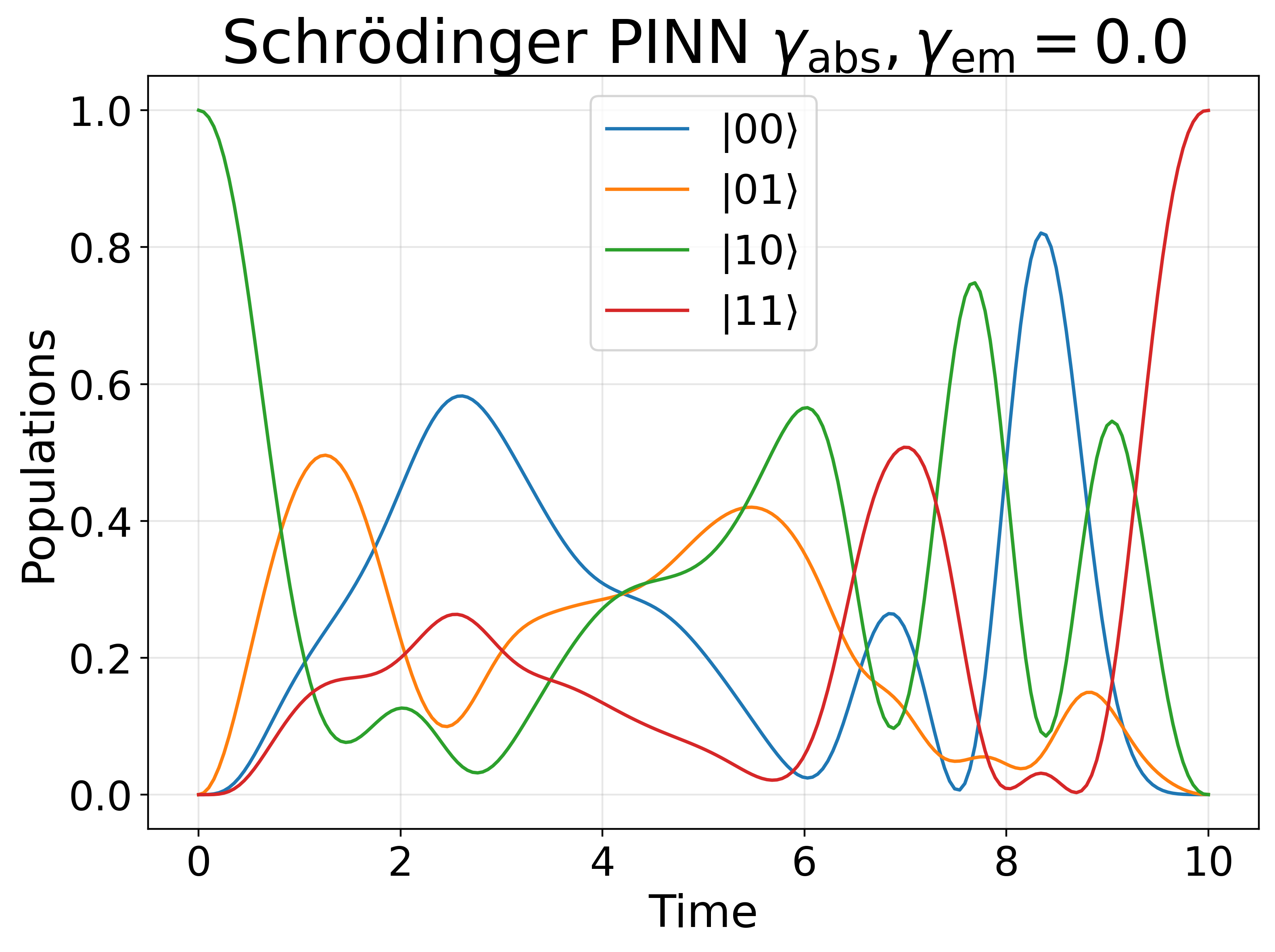}
  \end{subfigure}
  \begin{subfigure}[b]{0.45\linewidth}
      \centering\includegraphics[width=\linewidth]{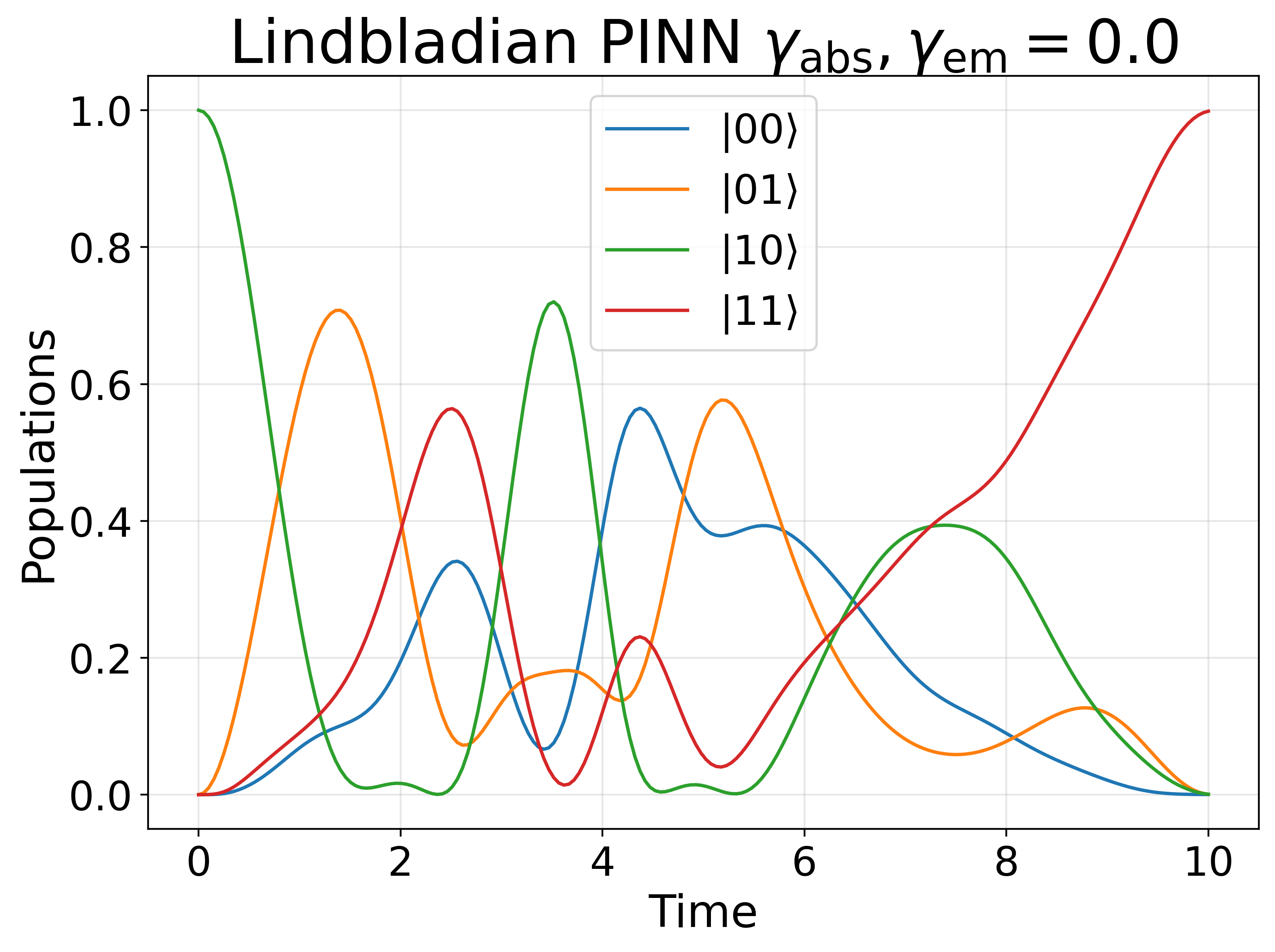}
  \end{subfigure}    
   \begin{subfigure}[b]{0.45\linewidth}
      \centering\includegraphics[width=\linewidth]{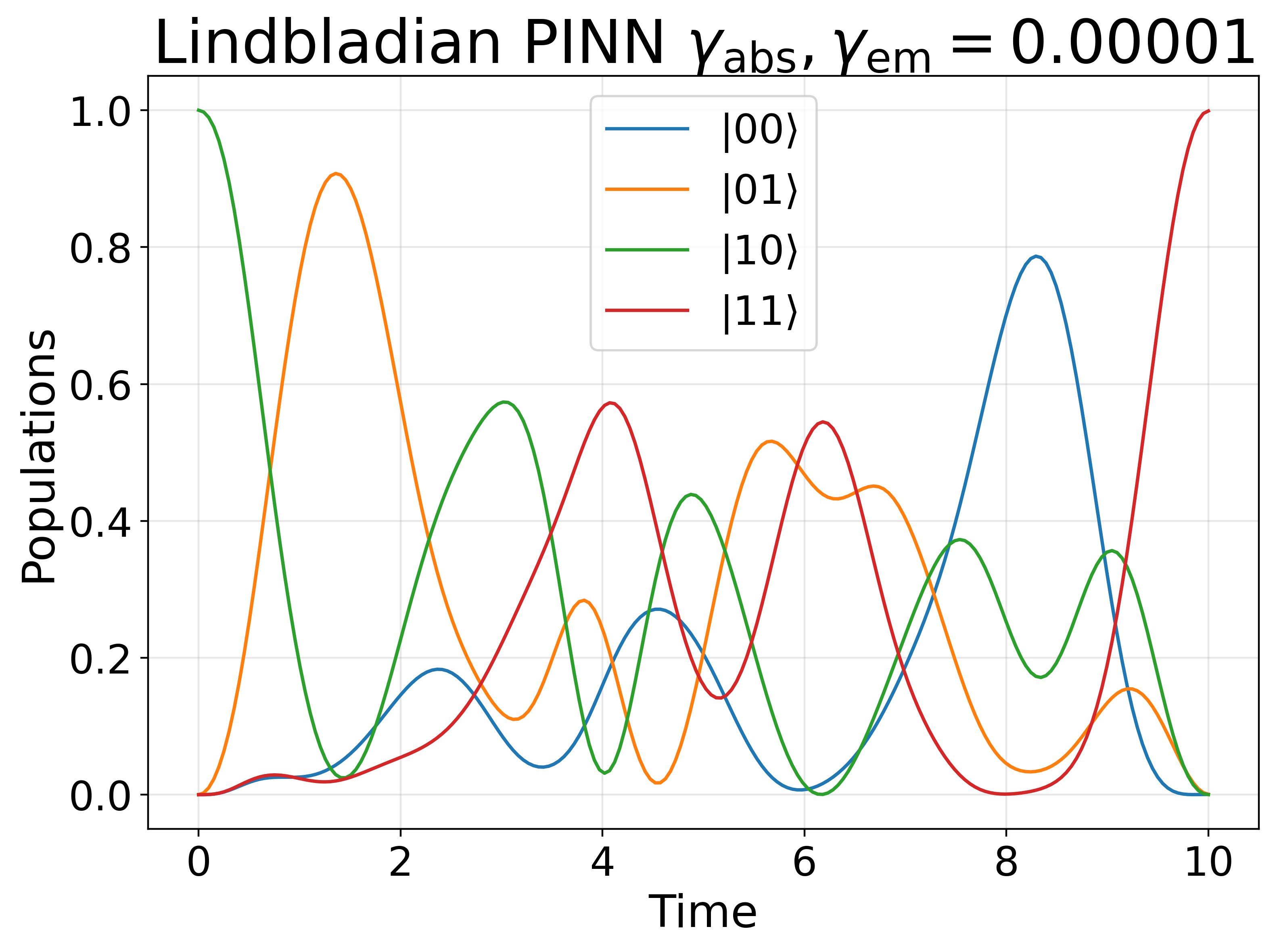}
  \end{subfigure}    
  \begin{subfigure}[b]{0.45\linewidth}   
        \centering\includegraphics[width=\linewidth]{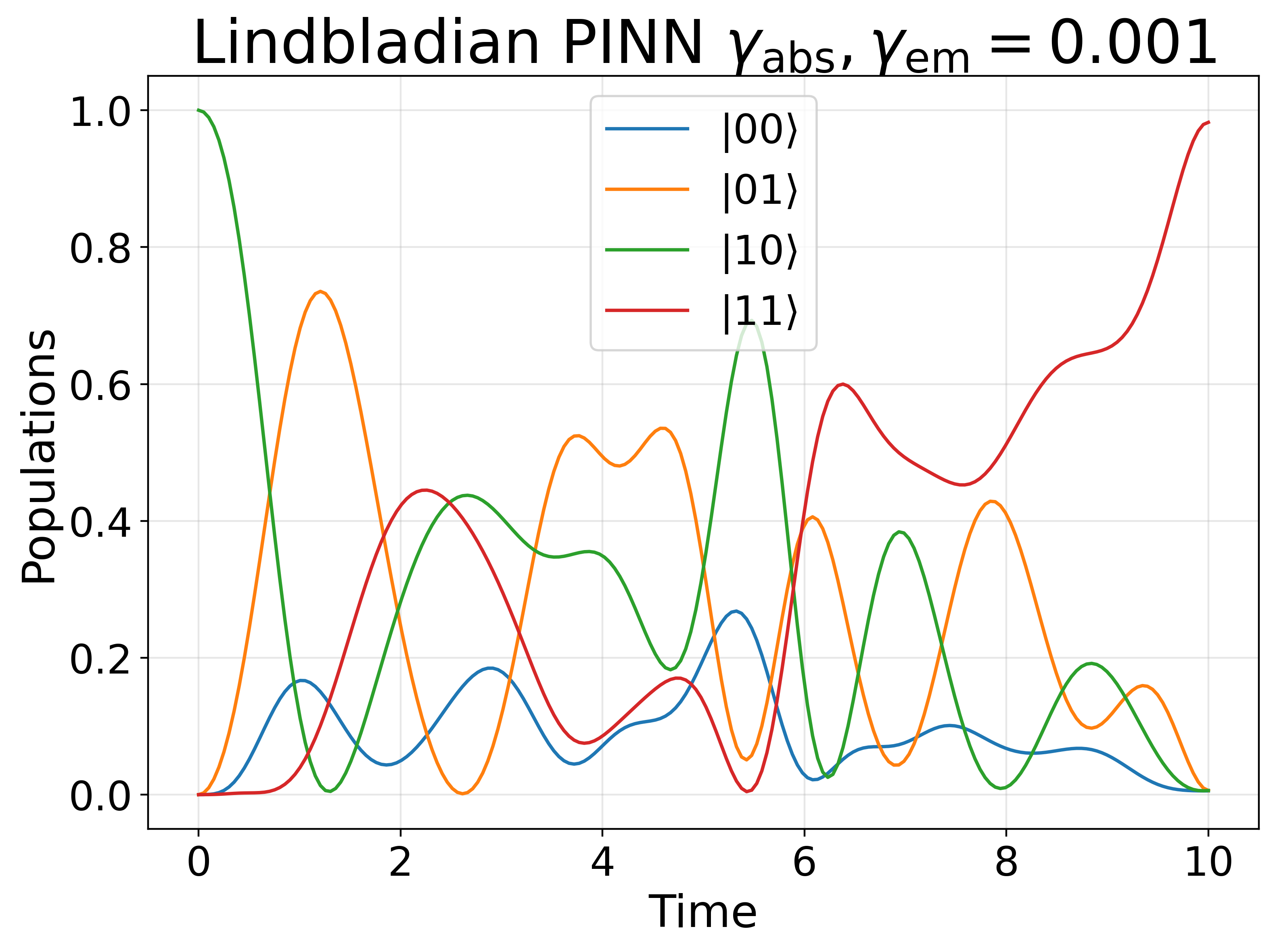}
  \end{subfigure}
  \begin{subfigure}[b]{0.45\linewidth}   
        \centering\includegraphics[width=\linewidth]{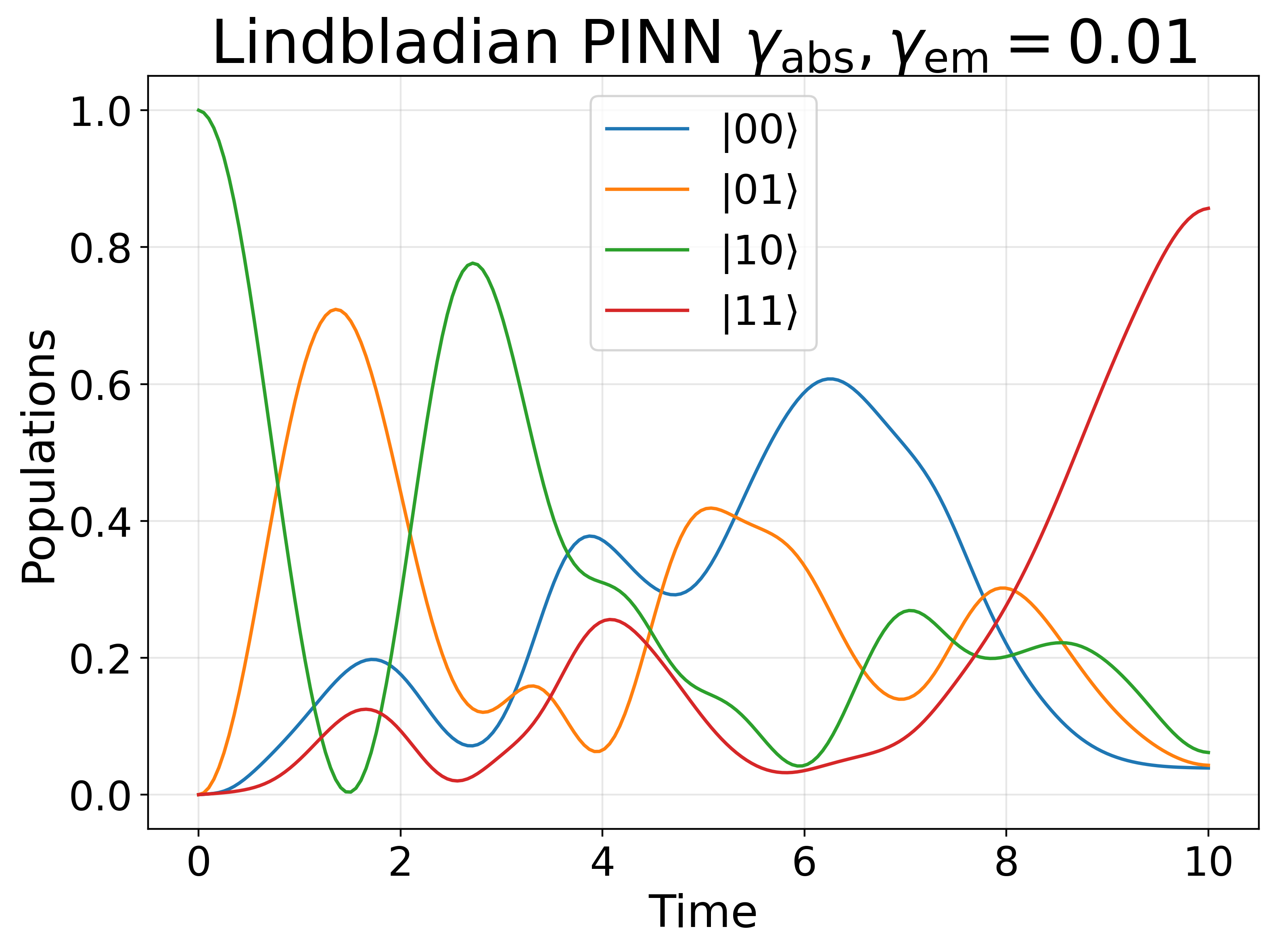}
  \end{subfigure}
  \begin{subfigure}[b]{0.45\linewidth}
      \centering\includegraphics[width=\linewidth]{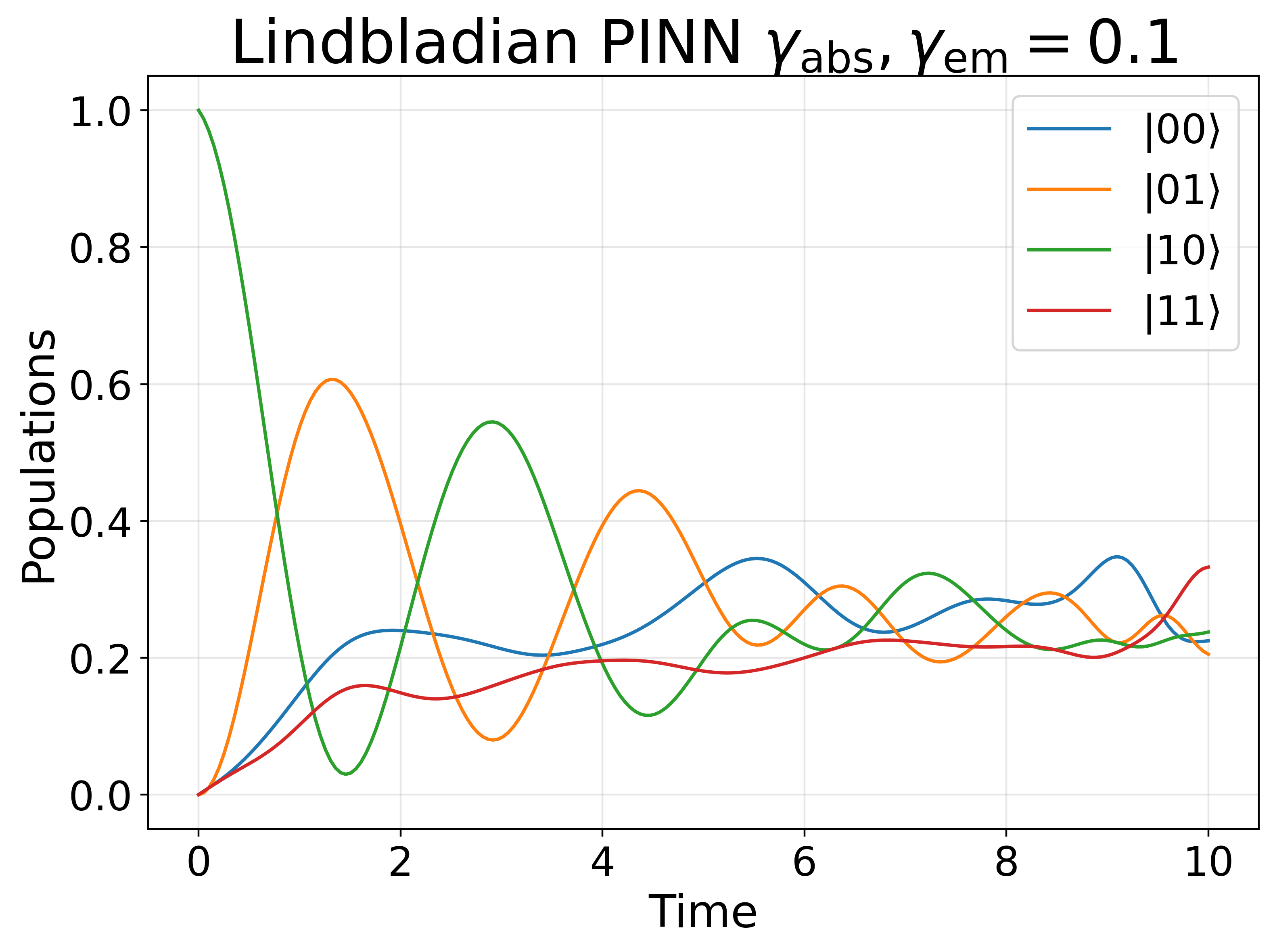}
  \end{subfigure} 
  \begin{subfigure}[b]{0.45\linewidth}
      \centering\includegraphics[width=\linewidth]{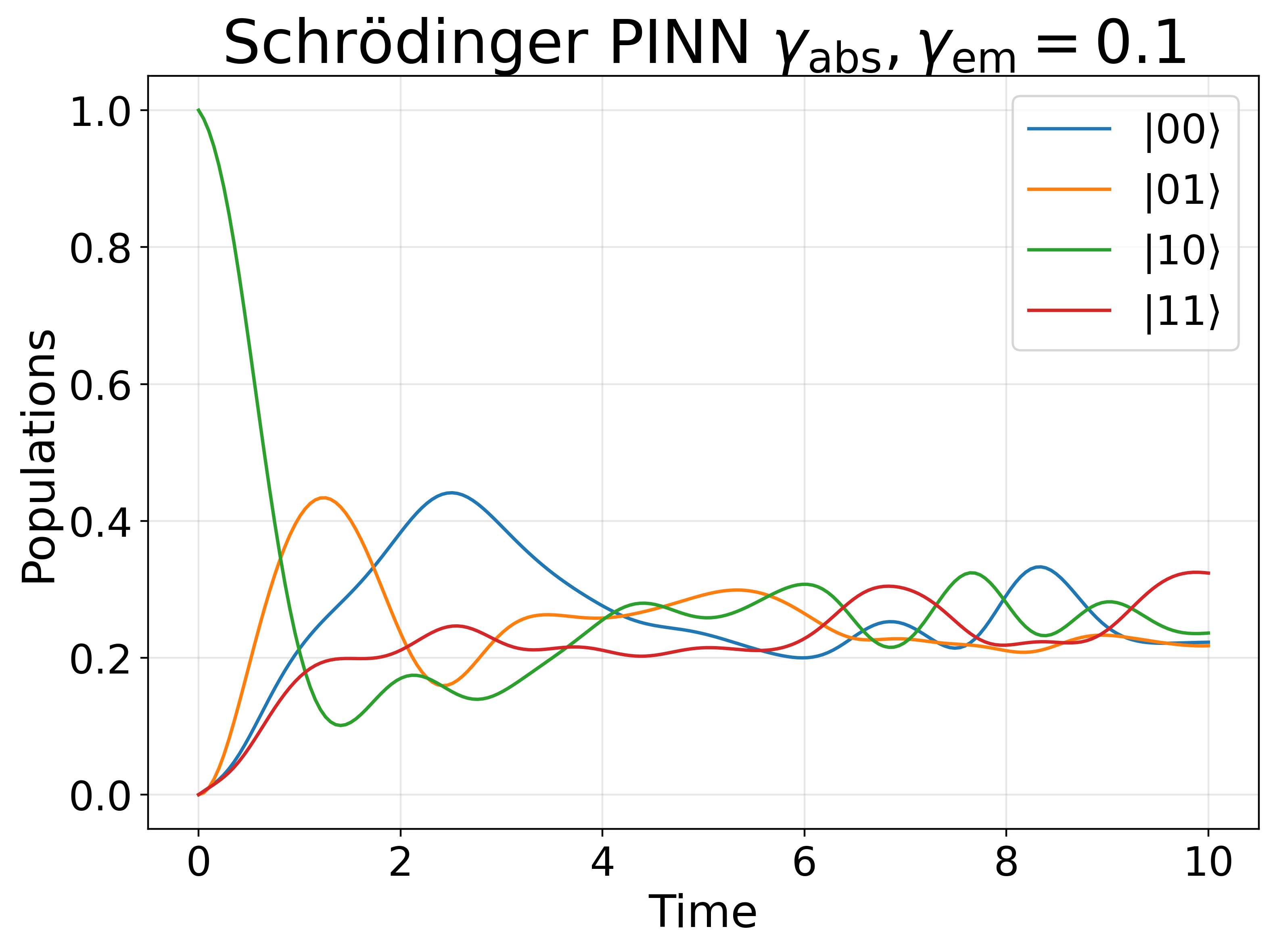}
  \end{subfigure} 
    \caption{Population graphs for different PINN models.} \label{fig:pop_graphs}   
  \end{figure}

In Fig.~\ref{fig:pop_graphs}, we show the time evolution of the computational basis state populations $(\ket{00}, \ket{01}, \ket{10}, \ket{11})$ under $U_{CNOT}$ gate learning for Lindblad PINN and Schrödinger PINN, using the same trained models that were utilized when generating Fig.~\ref{fig:training_loss} (and starting in ${x}_0 = \ket{10}$ state). We additionally include a Schrödinger PINN population graph, evaluated in a noisy environment, for comparison.

The first Schrödinger model, in a noise-free environment, and the first two Lindblad models successfully transfer population from the initial $\ket{10}$ state to the $\ket{11}$ state. The higher dissipation gets, the more evidently Lindblad PINN's struggle to perform a successful learning of the gate to populate the $\ket{11}$ state. In the last graph, we see that Schrödinger PINN fails in a very similar way to Lindblad PINN when evaluated in a noisy setting. 


\section{Conclusion}
In this work, we have introduced the relevance of ML applied to QOC. We explored the capabilities of PINN models and presented the details of our PINN with its architecture and the relevant ML choices, like weight initialization and activation function. We find that, for predicting pulses to implement quantum gates through a PINN, the use of $\sin(\cdot)$ activation function is preferable to other popular choices. The demonstrated custom variance scaling weight initialization choice was another discovered improvement for the learning behavior of our model. 

We created two types of generic PINN models for pulses to implement quantum gates that currently perform as well as the existing algorithm CRAB. We hypothesize that a PINN can potentially achieve full transferability of learning, where the inexpensive training can be run just once and the resulting model can then be generalized and evaluated on numerous gates. Such transferability of learning can be achieved through techniques like hypernetworks, or in this case, HyperPINNs \cite{de2021hyperpinn}. In that work, a physics-informed hypernetwork is trained over a family of system parameters. It outputs the main model that can estimate the solution for any gate configuration in the parameter range. If full transferability of learning is reached, then our model has a chance of achieving an advantage in computational cost over the existing algorithms like GRAPE and CRAB. 

When testing our PINN based on the Lindblad master equation and our PINN based on the Schrödinger equation in a noisy environment, both fail in similar ways. This suggests that a PINN needs more information on the noise rates and cannot learn pulses that are robust against noise by just training on a Lindblad equation. More work needs to be done to explore what kind of strategies would get us to more robust pulses. For now, it can be concluded that due to how similar both of our PINN versions perform (and since there was no particular supremacy of a Lindblad PINN trained with high noise rates, when evaluated in a noisy environment), it is appropriate just to use a Schrödinger version, since its training requires less computational cost.

We are planning on testing our PINN on superconducting qubit architectures like the tunable-coupling transmon \cite{stehlik2021tunable}, in addition to scaling our system to larger amounts of qubits, while keeping the cost reasonable (and seeing if it outperforms the existing techniques when scaled up). For further enhancement of our generic PINN, we are planning on adding important experimental constraints into the loss function to make our PINN-predicted pulses more suitable for use in a real experimental setup. Our ultimate goal is a low-cost, flexible, realistic, and generalizable model that automatically adapts to various quantum systems without retraining and outperforms existing algorithms.

\section{Acknowledgements}
We acknowledge funding from the AFOSR Young Investigator Program (AFOSR award FA9550-25-1-0150) and the UW-Madison's Letters and Science Honors Summer Research Apprenticeship.

\bibliography{ref}    
\end{document}